\documentclass{vgtc}                          

\ifpdf
  \pdfoutput=1\relax                   
  \usepackage{graphicx}                
  \DeclareGraphicsExtensions{.pdf,.png,.jpg,.jpeg} 
\else
  \usepackage{graphicx}                
  \DeclareGraphicsExtensions{.eps}     
\fi%

\graphicspath{{figures/}{pictures/}{images/}{./}} 

\usepackage{microtype}                 
\PassOptionsToPackage{warn}{textcomp}  
\usepackage{textcomp}                  
\usepackage{mathptmx}                  
\usepackage{times}                     
\usepackage{cite}                      
\usepackage{tabu}                      
\usepackage{booktabs}                  

\onlineid{0}

\vgtccategory{Research}

\vgtcinsertpkg

\title{An Open Platform for Research about Cognitive Load  in Virtual Reality}

\author{Olivier Augereau\thanks{e-mail: augereau@enib.fr}\\ %
        \parbox{1.4in}{\scriptsize \centering  Lab-STICC CNRS UMR6285 \\ ENIB, Brest, France\\} %
\and Gabriel Brocheton \thanks{e-mail:g7broche@enib.fr}\\ %
     \scriptsize ENIB, Brest, France %
\and Pedro Paulo Do Prado Neto" \thanks{e-mail:p1paulod@enib.fr}\\ %
     \scriptsize ENIB, Brest, France}

\abstract{The cognitive load can be used to assess if someone is struggling while performing a task. It can be used in many different situations such as in driving, piloting, studying, playing, working, etc. 
This information can help to design better systems and even to create interactive systems that can be aware of the user's cognitive load and adapt itself to the user.
We propose an open source platform that can be used for doing research about cognitive load in virtual reality (VR). Our platform can be used for stimulating cognitive load through several VR scenes and for analyzing cognitive load through objective and subjective measurements.
} 

\CCScatlist{ 
  \CCScat{H.5.1}{INFORMATION INTERFACES AND PRESENTATION}{Multimedia Information Systems}{Artificial, augmented, and virtual realities};
  \CCScat{J.4}{SOCIAL AND BEHAVIORAL SCIENCES}{Psychology}
  }

\begin{document}



\maketitle

\section{Introduction} 

Virtual Reality (VR) is a promising technology that the industry is starting to use more and more for professional training. Systems that are expensive, dangerous or not easily available can be modeled and inexperienced users can practice with them safely.
VR is also a useful tool for psychologists as they can reproduce specific experiments to analyze the user's mind and behavior.
Detecting automatically the mental workload in real time can help to design interactive systems that can adapt themselves to the user's needs. But it is still a research challenge, and open source platforms implementing cognitive load measurements and giving access to prefabricated scenes will hopefully help researchers.

In this paper we present a platform that can be used for stimulating and measuring cognitive load in virtual reality (VR). Our platform is open source and available on two GitLab repositories: one with a detailed documentation for developers\footnote{https://git.enib.fr/g7broche/vr-ppe} and another one ready to use for standard users\footnote{https://git.enib.fr/g7broche/vr-ppe-build}.

The following sections will present what kind of measurements can be done with the proposed platform and then the VR scenes that have been implemented in the platform to stimulate the cognitive load.

\section{Cognitive load measurement}
The cognitive load theory is based on human cognition and is characterized by the central assumption that the amount of knowledge acquisition depends on the efficiency
of the use of available cognitive resources in working memory.
A person’s cognitive load is the mental effort associated while performing a task or engaging in a learning process.
The cognitive load measurements are usually divided into two main categories: subjective measures and objective measures.

\subsection{Subjective measures}

Most research on cognitive load is using the subjective measurement to obtain the participant’s mental effort. 
It is subjective because is related to the user’s own perception of the amount of mental effort used while performing a task. The user is asked to answer some questions in order to estimate their level of cognitive load.

\begin{figure}[tb]
 \centering 
 \includegraphics[width=\columnwidth]{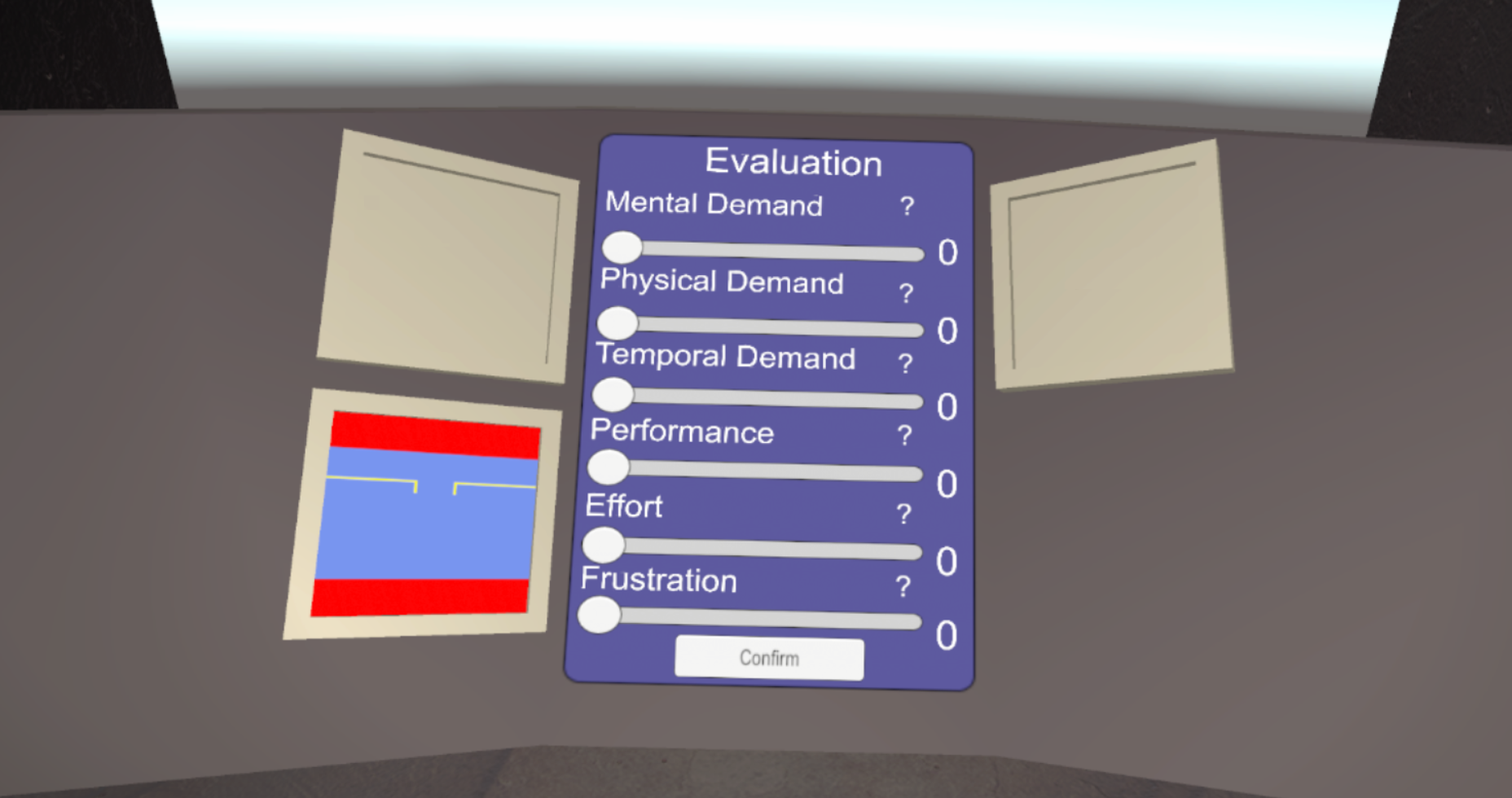}
 \caption{A visualization of the NASA-TLX questionnaire implemented in our VR platform.}
 \label{fig:nasatlx}
\end{figure}

One of the most famous subjective rating scales is the NASA Task Load Index (NASA-TLX) \cite{hart1988development}.
The NASA-TLX is a subjective rating scale, originally designed to analyze the mental and physical demands of aircraft pilots, with six different items : mental demand, physical demand, temporal demand, performance, effort, and frustration level. The \autoref{fig:nasatlx} shows the NASA-TLX implemented in our platform.

Another commonly used scale is the one proposed by Paas \cite{paas1992training}. It consists of one item for task difficulty and one item for mental effort on a 9-point Likert scale. 
These scales are placed for the participant to answer after each completed task.

The advantage of subjective measures is that the person who performed the task is the only one who can estimate the difficulty of the task faced, and the effort to complete it. 
However, there are technical measurement problems with regard to the objectivity, validity and reliability of the participants’ own responses.

\subsection{Objective measures}

The objective measurement is a way to obtain the user’s amount of mental effort by direct methods. 
These direct methods can be obtained through participant’s physiological information \cite{haapalainen2010psycho, ikehara2005assessing} or a dual-task performance \cite{park2015rhythm, brunken2002assessment}.

The analysis of physiological data obtained from the participant, such as brain activity, pupil movement and dilation, heart rate, electrodermal activity, etc. can be used to indicate the amount of mental effort made by the user. 
Physiological data can be an accurate indicator of the cognitive load, but the sensor's raw data usually need to be processed, features need to be extracted and sometimes machine learning models need to be trained before being able to predict the amount of mental effort based on these signals. Our platform exports all the physiological signals available from the VR headset for this purpose.

The dual-task performance method is another approach where the participants are engaged in two different tasks.
When the performance of the secondary task is reduced, it indicates an increase of the participant’s amount of mental effort, imposed by the primary task. Lower performances are typically reflected by a larger number of errors or a larger response time.

The values obtained by the dual-task paradigm are objectives, so analysis, and interpretation of data are measured independently of a self-report rating.
The problem of the dual-task methods is a possible interference between the primary and secondary tasks, when one of the tasks somehow prevents the execution of the other. Our platform logs the number of errors and the response time.

\section{Cognitive load stimulation}

Our VR platform was developed using the "Unity" game engine. Our platform is packaged as a single application composed of different scenes. 
The scenes are themselves composed of several phases (in our case two) each followed by a break where the NASA-TLX is displayed to the user. 
Each phase has its own set of tasks, all aimed at stimulating the user’s cognitive load. 

The tasks require users to use logical reasoning, concentration and to perform some simple mathematical calculations. They will be detailed in the following subsections.
During the tasks the user will hear audio beeps and is asked to press a button as soon as it is heard. This reaction time will be used as an objective measure of the cognitive load.

The game can be started by a unique executable file and the preferences of the game must be settled before launching the game, using a configuration file.
The configuration file allow to set some parameters such as: the user information, the scene that will be played, the execution order of the phases, the duration time of each phase and the pause, the use (or not) of NASA-TLX scale, the use (or not) of a tutorial, the repetition rate of audio beeps, etc.

All the data emitted by the VR headset sensors are recorded in a comma-separated values (CSV) file. In the same file, events that occur during the execution of scenes (phase start and end, task start and end, task completion or failure) are also recorded.
A text (TXT) file is also generated during the game. It will be filled with the user’s NASA-TLX answers, the average secondary task response time, and the average value of cognitive
load.
Both the CSV file and TXT file can be accessed at the end of the game. This information can then be used for future analysis.

We will now present the two main scenes available in the platform: the progressive task scene and the dual-task scene.

\subsection{Progressive task}

The aim of the progressive task scene is to stimulate the user's cognitive load progressively through a sequence of increasing hard phases. 
We implemented a mathematical calculation task where the user needs to point at a numeric keyboard to enter the result of the displayed calculation.
In the first phase, only arithmetic sums between numbers of maximum two digits will be displayed while during the second phase the sums can be between numbers of three digits.

\subsection{Dual-Task}

In this scene the user starts with a single task during the first phase, and then has to perform two different tasks at the same time in the second phase. 
During the first phase, the user needs to keep a line that is moving inside a defined interval by pressing a controller's button.  
During the second phase, the user keeps doing the same task plus another one. In our case he will have to perform a simple calculation (as described in the previous subsection).

\begin{figure}[tb]
 \centering 
 \includegraphics[width=\columnwidth]{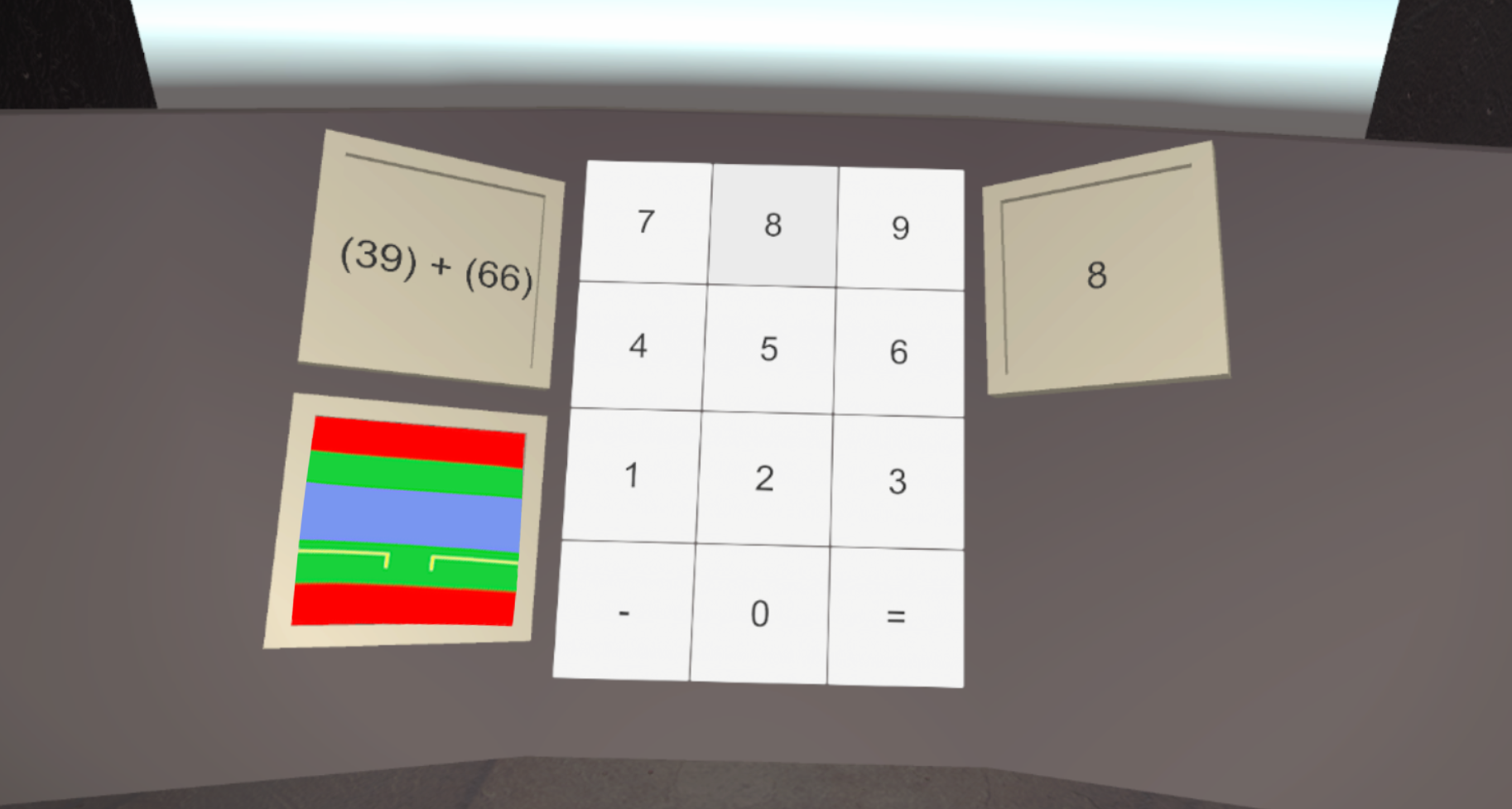}
 \caption{A visualization of the dual-task scenario. The user needs to do a mental calculation and to maintain a line in a specific interval at the same time. The application logs the user' errors, physiological signals, NASA-TLX answers, and the response time to audio beeps.}
 \label{fig:doubleTaskScenario}
\end{figure}

\section{Conclusion and Future Work}

We presented a VR platform that can be used for stimulating and measuring the cognitive load in various ways.

For now, our platform is only compatible with the VR headset "HP reverb G2 Omnicept"\footnote{https://www.hp.com/us-en/vr/reverb-g2-vr-headset-omnicept-edition.html}, but further work will be done to integrate other headsets such as the "HTC Vive Pro Eye".

The HP reverb G2 Omnicept is build in with a Tobii eye tracker, a photoplethysmogram sensor (it detects the blood volume changes such as the heart beat), a face camera located on the bottom of the headset pointing at user's mouth, and an inertial measurement unit including 3-axis accelerometers and 3-axis gyroscopes.
The headset also has a machine learning model that can predict the user's cognitive load as a continuous value ranging from 0.0 to 1.0.

One of the purposes of building this platform is to check if the cognitive load value given by the headset is usable and to which extent.
In order to do this we will start an experiment and collect data of some users to see if the value is correlated to other cognitive load measurements available in our platform such as the reaction time, the NASA-TLX of the dual-task performances.

\acknowledgments{
This work has been done in the European Center of Virtual Reality (CERV)\footnote{cerv.enib.fr}, a research platform of ENIB. We would like to thank the support of other CERV members and especially Nathalie Le Bigot for her expertise in psychology.}

\bibliographystyle{abbrv-doi}

\bibliography{template}

\begin{thebibliography}{1}

\bibitem{brunken2002assessment}
R.~Br{\"u}nken, S.~Steinbacher, J.~L. Plass, and D.~Leutner.
\newblock Assessment of cognitive load in multimedia learning using dual-task
  methodology.
\newblock {\em Experimental psychology}, 49(2):109, 2002.

\bibitem{haapalainen2010psycho}
E.~Haapalainen, S.~Kim, J.~F. Forlizzi, and A.~K. Dey.
\newblock Psycho-physiological measures for assessing cognitive load.
\newblock In {\em Proceedings of the 12th ACM international conference on
  Ubiquitous computing}, pp. 301--310, 2010.

\bibitem{hart1988development}
S.~G. Hart and L.~E. Staveland.
\newblock Development of nasa-tlx (task load index): Results of empirical and
  theoretical research.
\newblock In {\em Advances in psychology}, vol.~52, pp. 139--183. Elsevier,
  1988.

\bibitem{ikehara2005assessing}
C.~S. Ikehara and M.~E. Crosby.
\newblock Assessing cognitive load with physiological sensors.
\newblock In {\em Proceedings of the 38th annual hawaii international
  conference on system sciences}, pp. 295a--295a. IEEE, 2005.

\bibitem{paas1992training}
F.~G. Paas.
\newblock Training strategies for attaining transfer of problem-solving skill
  in statistics: a cognitive-load approach.
\newblock {\em Journal of educational psychology}, 84(4):429, 1992.

\bibitem{park2015rhythm}
B.~Park and R.~Bruenken.
\newblock The rhythm method: A new method for measuring cognitive load—an
  experimental dual-task study.
\newblock {\em Applied Cognitive Psychology}, 29(2):232--243, 2015.

\end{thebibliography}
\end{document}